\documentclass[a4paper,fleqn,usenatbib]{mnras}
\usepackage{times,txfonts}
\usepackage[T1]{fontenc}
\usepackage{natbib}
\DeclareRobustCommand{\VAN}[3]{#2}
\let\VANthebibliography\thebibliography
\def\thebibliography{\DeclareRobustCommand{\VAN}[3]{##3}\VANthebibliography}
\usepackage{graphicx}	
\usepackage{amsmath}	
\usepackage{amssymb}	

\title[Hot DAV white dwarf stars]{Pulsation in the white dwarf HE~1017--1352: confirmation of the class of hot DAV stars}

\author[Romero et al.]{Alejandra D. Romero$^{1},$\thanks{E-mail: alejandra.romero@ufrgs.br}
L. Antunes Amaral$^{1}$,
S. O. Kepler$^{1}$,
L. Fraga$^{2}$, 
 D. Kurtz$^{3,4}$ and
 \newauthor H. Shibahashi$^{5}$ 
\\
$^{1}$Physics Institute, Universidade Federal do Rio Grande do Sul, 91501--900 Porto--Alegre, RS, Brazil\\
$^{2}$Laborat\'orio Nacional de Astrof\'isica LNA/MCTIC, 37504-364 Itajub\'a, MG, Brazil\\
$^{3}$Department of Physics, North West University, Dr Albert Luthuli Drive, Mahikeng 2735, South Africa\\
$^{4}$Jeremiah Horrocks Institute, University of Central Lancashire, Preston PR1 2HE, UK\\
$^{5}$Department of Astronomy, School of Science, University of Tokyo, Tokyo 113--0033, Japan
}

\date{Accepted XXX. Received YYY; in original form ZZZ}

\pubyear{2020}

\begin{document}
\label{firstpage}
\pagerange{\pageref{firstpage}--\pageref{lastpage}}
\maketitle

\begin{abstract} 
We report the detection of periodic variations on the 
$T_\mathrm{eff}\simeq 32\,000$~K DA white dwarf star HE~1017--1352. We obtained time series photometry using the 4.1-m SOAR telescope on three separate nights for a total of 16.8~h. From the frequency analysis we found four periods of 605~s, 556~s, 508~s and 869~s with significant amplitudes above the 1/1000 false alarm probability detection limit. The detected modes are compatible with low harmonic degree g-mode non-radial pulsations with radial order higher than $\sim 9$. This detection confirms the pulsation nature of HE~1017--1352 and thus the existence of the new pulsating class of hot DA white dwarf stars. In addition, we detect a long period of 1.52 h, compatible with a rotation period of DA white dwarf stars. 
\end{abstract}


\section{Introduction}

Pulsating white dwarf stars can be found in several temperature ranges along the white dwarf evolutionary cooling tracks.  The excitation mechanisms are a combination of an opacity bump due to partial ionisation of the main element in the outer layers ($\kappa$--mechanism), and the effect of a small value of the adiabatic exponent $\Gamma_3 -1$ in the ionisation zone, preventing a large increase in temperature upon compression ($\gamma$-mechanism). Since ionisation occurs at different temperatures for different elements, pulsation instabilities are expected at different effective temperatures. For instance, the excitation mechanism for helium atmosphere white dwarfs (DB) is consistent with the $\kappa-\gamma$ mechanism due to the first helium ionisation (He\,I -- He\,II), as predicted by \citet{1982ApJ...252L..65W}.  The pulsating DB white dwarf stars are found in the temperature range of $25\,000\,{\rm K} \lesssim T_{\rm eff} \lesssim 30\,000\,{\rm K}$. Hydrogen atmosphere white dwarf stars (DA) near the blue edge of the DB instability strip could develop pulsation, if the hydrogen layer is thin enough. This hypothesis was proposed by \citet{2005EAS....17..143S,2007AIPC..948...35S} to predict the existence of pulsation in `hot DA' white dwarf stars, with effective temperatures around $\sim 30\,000$\,K.\footnote{\citet{1982ApJ...252L..65W} found that the He partial ionisation could also excite pulsations in DA white dwarfs with $M_H/M_* < 10^{-10}$ and $T_{\rm eff} \sim 19\, 000$ K.} 

\citet{2005EAS....17..143S,2007AIPC..948...35S} found that around $30\,000$\,K the atmosphere of a white dwarf is superadiabatic and convectively stabilised by a chemical composition gradient ($\mu$--gradient). Gravitational settling is very efficient for white dwarfs, leading to a stratified structure, where the lighter, cooler hydrogen is floating on top of the heavier helium layer. Under this condition he found that the radiative heat exchange leads to an asymmetry in g-mode oscillatory motion such that the oscillating elements overshoot their equilibrium positions with increasing velocity. A linear local stability analysis based on the dispersion relation led him to predict that g~modes should be excited in DA stars at the blue edge of the DBV instability strip, pulsating in higher degree modes, but with some $\ell <3$ modes excited. In the models, this could only occur for thin hydrogen envelopes of about $10^{-12}$\,M$_\odot$. However, the mass of the hydrogen envelope must be larger than about $10^{-14}$\,M$_\odot$, otherwise it will be diluted into the more massive helium layer due to convective mixing, and the star will no longer be a DA white dwarf for $T_{\rm eff}\sim 30\,000$\,K \citep{1987fbs..conf..319F}. 

Following the predictions of \citet{2005EAS....17..143S,2007AIPC..948...35S},  \citet{Kurtz2008} performed a search for pulsations in a sample of 7 DA white dwarf stars with spectroscopic effective temperatures in the effective temperature range $31200 - 29500$\,K. They reported two possible variable hot DAV stars: SDSS J010415.99$+$144857.4 and SDSS J023520.02--093465.3, with periods of 750~s and 159~s, respectively. Later, \citet{Kurtz2013} found a third candidate, HE~1017--1352 (WD~1017--138), showing a period of 615~s, based on two independent 2.3-h and 2.1-h photometric time series obtained with the 4.2-m William Herschel Telescope (WHT). 

\begin{figure}
\includegraphics[width=\columnwidth]{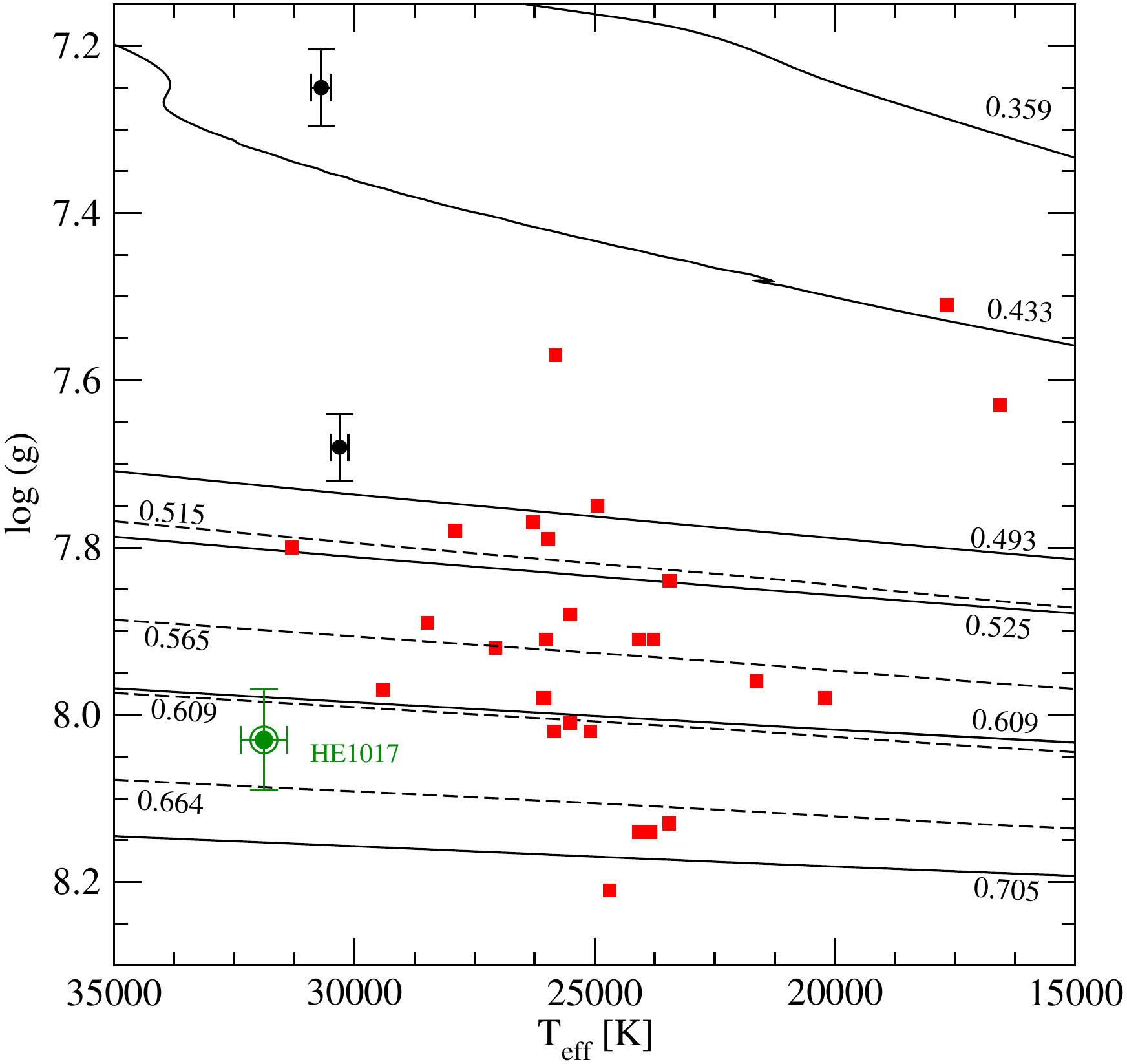}
\caption{The position of the three hot DAV candidates and DBV white dwarf stars in the $T_{\rm eff}-\log g$ plane. Hot DAV candidates from \citet{Kurtz2008} are depicted with black circles, while HE~1017--1352 is depicted with a green circle. Known DBV stars are depicted with red squares. Some cooling tracks with thin hydrogen envelope \citep[full line,][]{2019MNRAS.484.2711R} and helium pure atmospheres \citep[dashed line,][]{2009ApJ...704.1605A} are included for masses from 0.493\,M$_\odot$ to 0.705\,M$_\odot$. Two low mass sequences with thick hydrogen envelopes, with stellar masses of 0.359\,M$_\odot$ and 0.433\,M$_\odot$, are also included \citep{2019cwdb.confE..13R}. Atmospheric parameters for HE~1017$-$1352 are taken from \citet{2011ApJ...743..138G}, and for the other two hot DAV candidates are taken from \citet{2016MNRAS.455.3413K}.}
\label{Teff-logg}
\end{figure}

The Hot DAV stars presented by \citet{Kurtz2008,Kurtz2013} are depicted in Fig.\,\ref{Teff-logg} in the $T_{\rm eff}-\log g$ plane. The target of this study, HE~1017--1352,  is depicted by a green circle, with atmospheric parameters of $T_{\rm eff} = 32000 \pm 500$~K and $\log g = 8.03 \pm 0.06$, taken from \citet{2011ApJ...743..138G}. The two hot DAV candidates presented in \citet{Kurtz2008} are depicted with black circles (see \citealt{2016MNRAS.455.3413K} for details). For comparison, we plot 27 of the 46 known pulsating DB white dwarf stars with red squares  \citep{2015A&A...583A..86K,Zach2018}. Theoretical cooling sequences for thin hydrogen and pure helium envelope white dwarf models are included for stellar masses from 0.493\,M$_\odot$ to  0.705\,M$_\odot$, along with low mass  hydrogen envelope sequences.

Even after the works of \citet{Kurtz2008, Kurtz2013}, confirmation of variability in hot DA white dwarf stars is still needed to establish the existence of a new class of pulsating white dwarf stars. In this work, we perform follow-up observations on HE~1017--1352 with the 4.1-m Southern Astrophysical Research (SOAR) Telescope. We present our data analysis and results in Section~\ref{obs}. In addition, we present data from the TESS satellite and compared both with the original data from \citet{Kurtz2013}. Concluding remarks are presented in Section \ref{final}.

\section{The suspect: HE~1017--1352}
\label{obs}

HE~1017--1352 ($\alpha_{2000}$ = 10:19:52.36, $\delta_{2000} = -$14:07:34.26) is a $V=14.6$ DA white dwarf star with an effective temperature close to the blue edge of the DBV instability strip, as shown in Fig.\,\ref{Teff-logg}. \citet{2011ApJ...743..138G} found atmospheric parameters of $T_{\rm eff} = 32000 \pm 500$\,K and $\log g = 8.03 \pm 0.06$ (cgs units), using atmosphere models with a mixing length of $0.8\,H_p$, where $H_p$ denotes the pressure scale height, while \citet{2009A&A...505..441K} found values of $T_{\rm eff} = 31800\pm 30$\,K and $\log g = 7.840\pm 0.006$, using models with $0.6\,H_p$. Since the models do not have significant surface convection, the value of $\alpha$ is not dominant. Note that the
quoted uncertainties correspond to the statistical uncertainties of the
fitting procedure and are probably underestimated. This two determinations are in agreement within 0.2~$\sigma$ in $T_{\rm eff}$ and 3.2~$\sigma$ in $\log g$, but most importantly, they agree that HE~1017--1352 is a hot DA white dwarf star. 

From the spectroscopic parameters we computed the stellar mass using the white dwarf evolutionary sequences from \citet{2019MNRAS.484.2711R}. For theoretical sequences with canonical hydrogen envelopes -- those obtained from single stellar evolution computations -- we obtained a stellar mass of $0.661 \pm 0.030$\,M$_\odot$ considering the spectroscopic values from \citet{2011ApJ...743..138G} and $0.566 \pm 0.003$\,M$_\odot$ for the those obtained by  \citet{2009A&A...505..441K}.  The stellar mass is reduced to $0.636 \pm 0.031$\,M$_\odot$ and $0.541 \pm 0.003$\,M$_\odot$, respectively, if we consider theoretical sequences with thin hydrogen envelopes ($M_H \sim 10^{-9.5} M_*$).
We can also estimate the stellar mass from the parallax and magnitudes from the Gaia mission data release 2, independently from spectroscopy \citep[see][for details]{2019MNRAS.490.1803R}. We employ thin hydrogen atmosphere models from the Montreal group\footnote{\url{http://www.astro.umontreal.ca/~bergeron/CoolingModels/} }\citep{1995PASP..107.1047B} for Gaia magnitudes to transform absolute magnitude and colour into stellar mass and effective temperature. For HE~1017--1352 we find that $\pi = 9.34\pm 0.14$\,mas, $G = 14.559$ and $G_{b_p}-G_{r_p} = -0.439$, leads to a stellar mass of $0.421\pm 0.008$\,M$_\odot$, $22 - 33$~per~cent lower than the value obtained from spectroscopy. The photometric effective temperature in this case is around 31200\,K.
 
\subsection{SOAR data: frequency analysis and discussion}

For the observations of HE~1017--1352, we employed the Goodman spectrograph in image mode on the 4.1-m Southern Astrophysical Research (SOAR) Telescope. We used the CCD binned 2$\times$2 and a Region of Interest (ROI) of 800$\times$800 pixels, yielding a plate scale of 0.30 arcsec/pixel and a field of view of 4$\times$4 arcmin. This setup yielded a readout time of 7\,s. All observations were obtained with a red blocking filter S8612 (transmitting $\lambda = 330-620$ nm) to decrease sky contamination. We observed the target for three nights in the beginning of 2020, for a total of 16.76\,h. The journal of observations is presented in Table~\ref{journal}. 

\begin{table}
	\centering
	\caption{Journal of observations for HE~1017--1352 with the SOAR telescope. $\Delta t$ is the length of each observing run and $t_{\rm exp}$ is the exposure time of each exposure.}
	\label{journal}
	\begin{tabular}{ccc} 
		\hline
 Run start (UT)  &  $t_{\rm exp}$ (s) &  $\Delta t$ (h) \\
\hline
2020-01-22 04:44:23.35 & 5 & 3.76 \\
2020-03-07 01:16:48.89 & 5 & 6.07 \\
2020-03-08 01:08:33.73 & 5 & 6.93 \\
\hline 
	\end{tabular}
\end{table}

\begin{figure*}
\includegraphics[width=\textwidth]{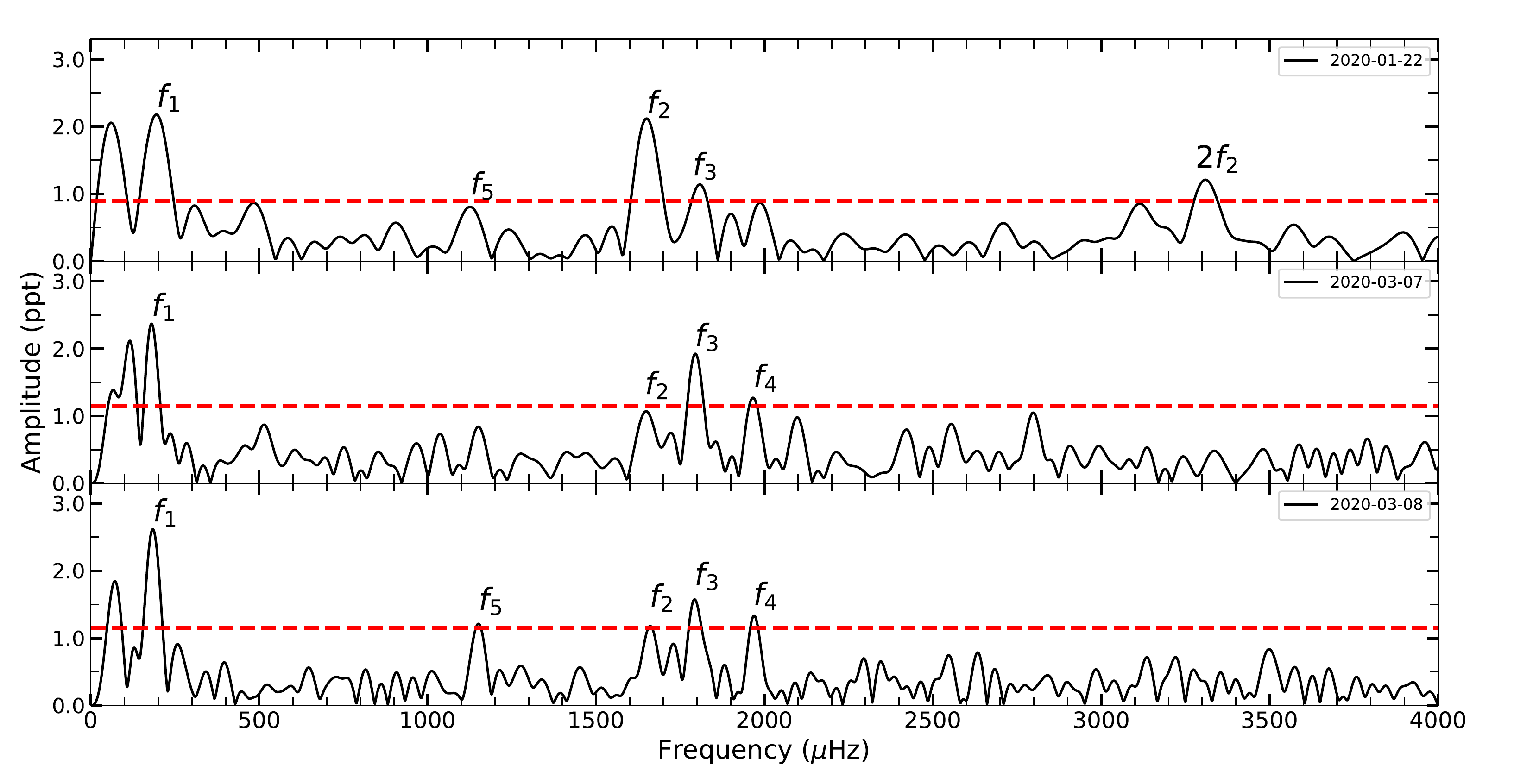}
\caption{Fourier Transform for HE~1017--1352 for the SOAR observations for each of the three nights listed in Table~\ref{journal}. The 1/1000 False Alarm Probability detection limit was computed using random shuffling of the data. 
The identification for each peak is indicated (see Table \ref{tab-HE-soar}). The FT corresponding to the concatenation of the three nights is depicted in the middle panel of Fig. \ref{HE17}. }
\label{HE17SOAR}
\end{figure*}

\begin{table*}
	\centering
	\caption{Detected frequencies for HE~1710--1352 from the observations performed with SOAR. We list the frequency, amplitude and period for each peak in the FT depicted in Fig.\,\ref{HE17SOAR} for each night. The quoted uncertainties correspond to the internal uncertainties. The value for the detection limit with 99.9~per~cent confidence is indicated for each night in the first row. The last column shows the identification of the mode. The list of frequencies for the complete light curve, considering the concatenation of the three nights, is presented in Table~\ref{tab-HE}. }
	\label{tab-HE-soar}
	\begin{tabular}{ccccccccccl} 
		\hline
Freq ($\mu$Hz) & Amp (ppt) & $\Pi$ (s) & Freq ($\mu$Hz) & Amp (ppt) & $\Pi$ (s) & Freq ($\mu$Hz) & Amp (ppt) & $\Pi$ (s) & ID\\
		\hline
\multicolumn{3}{c}{{\bf 2020-01-22} (0.9 ppt)}  &  \multicolumn{3}{c}{{\bf 2020-03-07} (1.1 ppt)} & \multicolumn{3}{c}{{\bf 2020-03-08} (1.2 ppt)} & \\
\hline
$195 \pm 5$ & $2.2 \pm 0.2$ & $5107 \pm 130$   & $180 \pm 3$ & $2.4 \pm 0.4$ & $5539 \pm 92$ & $184 \pm 2$ & $2.6 \pm 0.3$ & $5424 \pm 59$ & $f_1$\\
$1651 \pm 4$ & $2.1 \pm 0.2$ & $605 \pm 1$ & $1645 \pm 6$ & $1.2 \pm 0.2$ & $607 \pm 2$ & $1659 \pm 4$ & $1.3 \pm 0.2$ & $602 \pm 1$ & $f_2$\\
$1787 \pm 7$ & $1.0 \pm 0.2$ & $559 \pm 2$ & $1793 \pm 4$ & $1.9 \pm 0.3$ & $557 \pm 1$ & $1793 \pm 3$ & $1.6 \pm 0.3$ & $557.6 \pm 0.9$ & $f_3$ \\
$3309 \pm 5$ & $1.3 \pm 0.2$ & $302.1 \pm 0.5$ & $\cdots$ & $\cdots$ & $\cdots$ & $\cdots$ & $\cdots$ & $\cdots$ & $2f_2$ \\
$\cdots$ & $\cdots$ & $\cdots$ & $1967 \pm 7$ & $1.2 \pm 0.3$ & $508 \pm 2$ & $1967 \pm 4$ & $1.3 \pm 0.2$ & $508 \pm 1$ & $f_4$ \\
$1126 \pm 9$ & $0.8 \pm 0.1$ & $887 \pm 7$ & $\cdots$ & $\cdots$ & $\cdots$ & $1152 \pm 5$ & $1.2 \pm 0.2$ & $867 \pm 4$ & {\it $f_5$} \\
\hline
	\end{tabular}
\end{table*}

We reduced the data with the IRAF software, and performed aperture photometry with the DAOPHOT task. We used differential photometry to minimise effects of sky and transparency fluctuations. We tested all the comparison stars in the field; from the three brightest, we used the light curve with the best S/N. To look for periodicities in the light curves, we calculated a discrete Fourier Transform (FT) using the {\tt Period04} software \citep{2004IAUS..224..786L}.

The FTs for each night are presented in Fig.\,\ref{HE17SOAR}, and the detected frequencies are listed in Table~\ref{tab-HE-soar}. The detection limit (dashed line) corresponds to the 1/1000 False Alarm Probability (FAP), where any peak with amplitude above this value has 0.1\% probability of being a false detection due to noise. The FAP is calculated by shuffling the fluxes in the light curve while keeping the same time sampling, and computing the FT of the randomized data. This procedure is repeated N/2 times, where N is the number of points in the light curve. For each run we compute the maximum amplitude of the FT. From the distribution of maxima we take the 0.999 percentile to be the detection limit. The internal uncertainties in frequency and amplitude where computed using a Monte Carlo method with 1000 simulations with {\tt Period04}, while uncertainties in the periods were obtained through error propagation.

For the three nights, a peak with long period at $\sim 5400$\,s ($\sim 1.5$\,h) is present with significant amplitude. This 1.5-h period is compatible with a rotation period of DA white dwarf stars, in particular for ZZ Ceti stars near the blue edge of the instability strip \citep{2017IJMPS..4560023K}. Note that this period is coherent over the three nights of observation. The peaks at low frequencies, for periods longer than 1.5-h, are a consequence of the interference of Earth atmosphere and they are also observed in the light curve of the comparison stars. In addition, our observations are not long enough to resolve such long periods.
The peak corresponding to a period of $\sim 606$\,s is dominant in the first night (top panel of Fig.\,\ref{HE17SOAR}), for which its harmonic is also present at 302.1\,s. For the second and third nights (middle and bottom panels), this mode decreased in amplitude, while the peak with period $\sim 557$\,s increased. As a consequence of the energy loss, the peak with 302.1\,s, corresponding to the harmonic of the $\sim 606$\,s peak, disappears. In addition, the peak corresponding to a period of $\sim 508$\,s increased its amplitude above the FAP (1/1000) detection limit.
As expected, we can better resolve the three frequencies in the second and third nights, since the observation time is longer than the beat period of $\sim 4$ h (see Table \ref{journal}). 

Finally, we computed the FT of the concatenation of the three nights of observations with the SOAR telescope, with 16.76~h of data, after correcting the timings to barycentric dynamical time. The result is presented in the middle panel of Fig.\,\ref{HE17}, and the list of detected frequencies, amplitudes and periods is presented in the first three columns of Table~\ref{tab-HE}. 
We find three modes with periods of 508.121\,s, 556.802\,s and 605.310\,s with significant amplitudes above the FAP(1/1000) detection limit. At lower frequencies, a low amplitude peak is also present in the FT, corresponding to a period of 869.358~s. These periods are consistent with low harmonic degree g~modes in white dwarf stars. 
If we consider that the periods of 508.121~s, 556.802~s and 605.310~s correspond to modes with consecutive radial order $k$, we can estimate the forward period spacing as $\Delta \Pi_k = \Pi_{k+1}-\Pi_{k} \sim 48-49$ s. This value is in agreement with the expected value of the asymptotic period spacing for $\sim 0.5-0.6$\,M$_{\odot}$ white dwarf stars \citep{1980ApJS...43..469T}.
In summary, we found four peaks in the FT for HE~1017--1352 consistent with pulsations. 

At low frequencies, the peak with the largest amplitude shows a period of 1.52~h, that we propose as a rotation period. However, it is possible that low frequency peaks to be sub--harmonics of shorter periods. While linear combinations tend to have smaller amplitudes than the parent modes, \citet{2015MNRAS.450.3015K} showed that it is theoretically possible for combination frequencies to have amplitudes greater than the base frequency amplitudes, and that this is observed in some $\gamma$ Dor and Slowly Pulsating B stars.
Nonetheless, the frequency corresponding to 1.52~h is not compatible with any linear combination of the three shorter detected periods, within the uncertainties.

We searched for a preliminary seismic solution, considering the periods obtained from the SOAR observations (see column 3 of Table~\ref{tab-HE}). We first computed adiabatic pulsations for a white dwarf model characterized by 0.632\,M$_{\odot}$, $T_{\rm eff}=31\,870$ K and $\log(M_H/M_{\odot}) = -12.6$ \citep[see][for details]{2012MNRAS.420.1462R}, following the spectroscopic parameters obtained from \citet{2011ApJ...743..138G}. For this model, we can fit the four observed periods 508.121, 556.802, 605.310 and 869.358~s, with $\ell=1$ theoretical modes with periods of 507.8, 556.1, 619.5 and 851.6~s, corresponding to radial number $k = 13, 14, 16$ and 24, respectively. In addition, we also considered a C/O core white dwarf model with 0.423\,M$_{\odot}$, $T_{\rm eff}=31\,200$ K and $\log(M_H/{\rm M}_{\odot}) = -12.6$, that matches the photometric stellar mass and effective temperature based on Gaia data. For this model, we can fit the four observed periods with theoretical periods of 518.9, 547.4, 601.5 and 882.8~s, corresponding to radial numbers $k= 10, 11, 12$ and 19. 

\subsection{Comparison with WHT and TESS data}

 \begin{figure*}
\includegraphics[width=\textwidth]{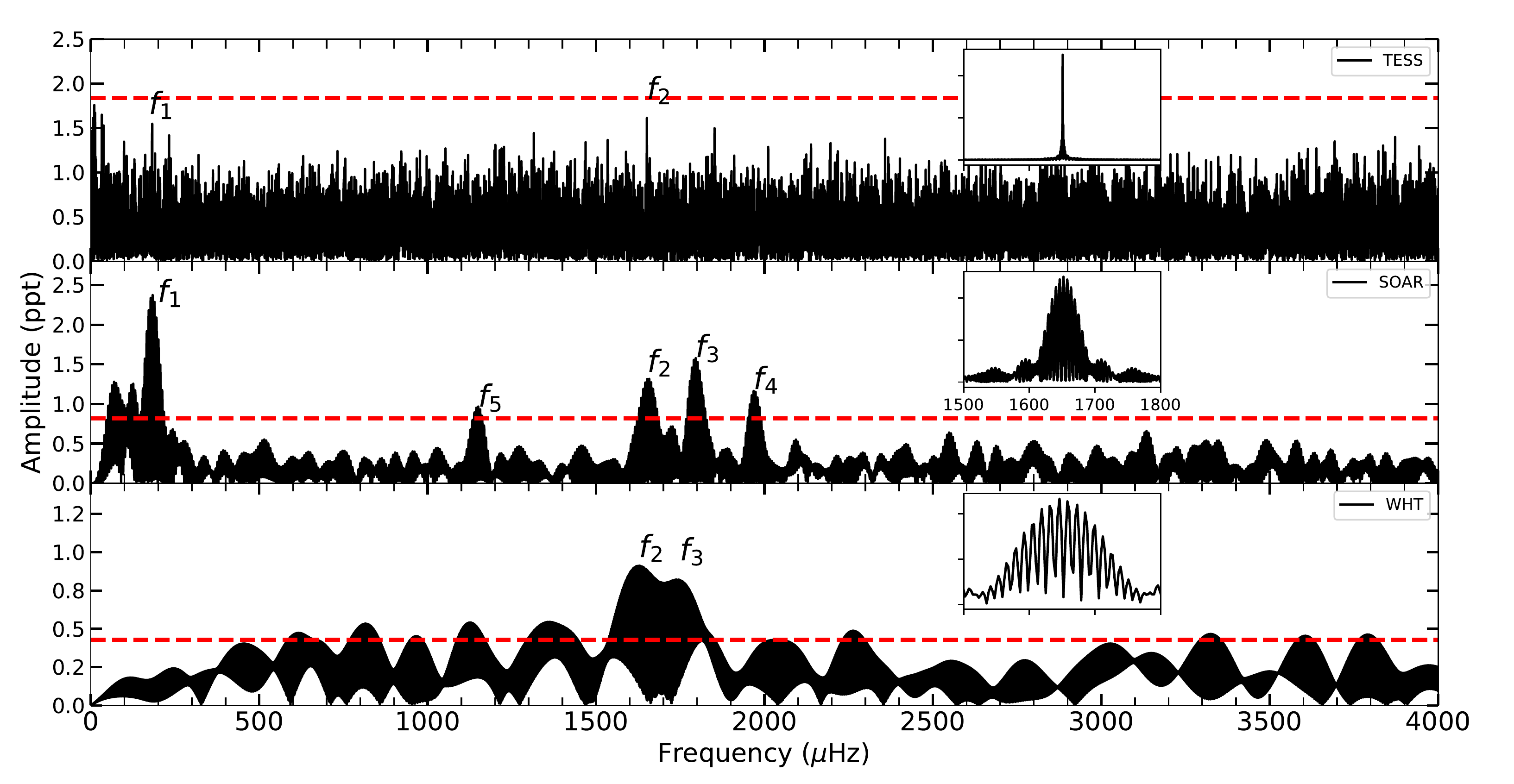}
\caption{Fourier Transform for HE~1017$-$1352 based on observations with three different instruments. Observations from TESS satellite (top) span for 24\,d in sector 9. The FT for SOAR telescope (middle) is computed from the concatenation of the three nights presented in the previous section. The FT for the WHT (bottom) was computed from the original data considering the concatenation of the two nights. The FAP (1/1000) detection limit (dashed line) was computed using random shuffling of the data. The identification for each peak is indicated (see Table \ref{tab-HE} for details). The spectral window for each case is depicted as an inset plot, with a full--width at half maximum (FWHM) of 1~$\mu$Hz for TESS, 68~$\mu$Hz for SOAR and 280~$\mu$Hz for WHT. The x-axis is in $\mu$Hz and all inset plots are in the same scale as in the middle inset plot.}
\label{HE17}
\end{figure*}

\begin{table*}
	\centering
	\caption{Frequency, amplitude and periods from observations with SOAR telescope (columns 1 to 3), TESS satellite (columns 4 to 6) and WHT (columns 7 to 9), corresponding to the FT shown in Figure \ref{HE17}. The quoted uncertainties correspond to the internal uncertainties. The value for the detection limit with 99.9~per~cent confidence is indicated in each case in the first row. The last column shows the identification of the mode. }
	\label{tab-HE}
  \scalebox{0.92}{%
	\begin{tabular}{cccccccccl} 
		\hline
Freq ($\mu$Hz) & Amp (ppt) & $\Pi$ (s) & Freq ($\mu$Hz) & Amp (ppt) & $\Pi$ (s) & Freq ($\mu$Hz) & Amp (ppt) & $\Pi$ (s) & ID \\
		\hline
\multicolumn{3}{c}{{\bf SOAR} (0.8 ppt)} & \multicolumn{3}{c}{{\bf TESS} (1.8 ppt)}&  \multicolumn{3}{c}{{\bf WHT} (0.429 ppt)} &  \\
$183.563 \pm 0.003$ & $2.4 \pm 0.1$ & $5447.73 \pm 0.09$ & $182.38 \pm 0.08$ & $1.5 \pm 0.3$ & $5483 \pm 2$ & $\cdots$ & $\cdots$ & $\cdots$ & $f_1$  \\
$1652.046 \pm 0.005$ & $1.4 \pm 0.1$  & $605.310 \pm 0.002$  & $1650.90 \pm 0.07$ & $1.6 \pm 0.3$ & $605.73 \pm 0.02$  & $1628.27 \pm 0.04$ & $0.92 \pm 0.07$    & $614.15 \pm 0.02$ &  $f_2$\\
$1795.970 \pm 0.004$ & $1.6 \pm 0.2$ & $556.802 \pm 0.002$ & $\cdots$ & $\cdots$ & $\cdots$ & $1748.87 \pm 0.05$  & $0.91 \pm 0.08$  & $571.80 \pm 0.02$ & $f_3$\\
$1968.034 \pm 0.007$ & $1.1 \pm 0.1$ & $508.121 \pm 0.002$ &  $\cdots$ & $\cdots$ & $\cdots$ &  $\cdots$ & $\cdots$ & $\cdots$ & $f_4$\\
$1150.274 \pm 0.008$ & $1.0 \pm 0.2$ & $869.358 \pm 0.006$ &  $\cdots$ & $\cdots$ & $\cdots$ &  $\cdots$ & $\cdots$ & $\cdots$ & $f_5$\\
\hline
\end{tabular}}
\end{table*} 

As mention before, photometric variability was reported for HE~1017--1352 from observations performed by \citet{Kurtz2013}. They observed this star using the WHT on two nights for a total of 4.4~h, and reported a period at 624\,s. 
For comparison purposes we re-computed the FT for the original data obtained by \citet{Kurtz2013}, shown in the bottom panel of Fig. \ref{HE17} after correcting the timings to barycentric dynamical time.  From our analysis, we found a period at 614\,s. The internal uncertainties for WHT data are underestimated, due to beating. Considering that the uncertainty related to the full width at half maximum (FWHM) of the peak in the WHT observations is 280\,$\mu$Hz, or 105\,s (see inset plot in bottom panel of Fig. \ref{HE17}), the period at 614\,s is in agreement with the peak at 605~s obtained from SOAR data. In addition, a second peak at 572\,s seems to be also present in the data from the WHT. However, due to the low resolution in frequency of the data it is not possible to separate it from the main peak. The list of frequencies, amplitudes and periods is presented in Table \ref{tab-HE}. Because the WHT data is composed by two chunks of $\sim$2-h long, separated by 3 days, there is a strong correlation between the peaks and a single FAP value is not reliable, leading to an underestimated  detection limit in this case. The additional peaks above the FAP(1/1000) limit that are present in the FT for the WHT data are part of the background noise and we only consider the two peaks listed in Table \ref{tab-HE} to be real periodicities.

HE 1017--1352 was also observed by the TESS satellite (TIC~307982318) in sector 9 for a 24-d run (2019-02-28T17:17:11.55 to 2019-03-25T23:30:40.318) with a 2 minute cadence. The light curve was downloaded from Mikulski Archive for Space Telescopes (MAST)\footnote{\url{ https://archive.stsci.edu/}}. We have used the PDCSAP flux that removes common instrumental trends giving a pre--search data conditioning (PDC). Additionally, we have performed a sigma--clipping for any data more than 5 sigma away from the mean value of the light curve. The FT for the TESS data is shown in the top panel of Figure \ref{HE17} (see also Table \ref{tab-HE}). As can be seen from this figure, the TESS data are not conclusive on the detection of pulsations since there is no peak with amplitude above its FAP (1/1000) detection limit. For white dwarf stars as hot as HE 1017--1352, the flux curve peaks at UV wavelengths. Unlike to the S816 filter used in the observations with SOAR, the TESS satellite bandpass is equivalent to a red filter, covering a wavelength range of $600 - 1000$\,nm, where the photometric amplitude for hot stars is very low compared to that at bluer wavelengths  \citep[e.g.][]{RKN}. For $T_{\rm eff} \sim 12\,000$~K DAV white dwarf stars it reduces their amplitude by about 63~per~cent \citep{2020arXiv200311481B}.
However, the largest peak occurs at 605.73~s. In addition, the peak at 5482.92~s, or 1.52~h, is present, in agreement with the long period detected in the SOAR data. 

\section{Conclusions}
\label{final}

We performed follow-up observations on the hot DA white dwarf star HE~1017--1352. This object was reported as variable by \citet{Kurtz2013}, following the theoretical predictions from \citet{2005EAS....17..143S,2007AIPC..948...35S}. We observed HE~1017--1352 with the SOAR telescope for a total of 16.76~h in three separate nights at the beginning of 2020. 
From the Fourier Transform corresponding to the combination of the three nights, we found three periods of 508.121\,s, 556.802\,s and 605.310\,s, with significant amplitudes above the detection limit. These periods are compatible with low harmonic degree g~modes, with radial orders $k\gtrsim 9$. A low amplitude mode with a period of 869.358\,s is present, also compatible with a low $\ell$ mode, but for radial order $k\gtrsim 17$. The period at 605.310\,s is in agreement with the peak at 614\,s obtained from the WHT data and reported by \citet{Kurtz2013}, within their uncertainties related to the FWHM of the peak of 280\,$\mu$Hz.
In addition, we detected a long period of 1.52~h, which is compatible with a rotation period. Evidence of this long period can also be found in the observations from the TESS satellite.

Our observations confirm the pulsational nature of HE~1017--1352 and thus the existence of the hot DAV class of pulsating stars. 
This opens the possibility to study the interior of these
objects through asterosesimology. In addition, the confirmation
of pulsations in hot DA stars is another evidence for the existence of white dwarfs with very thin hydrogen envelopes \citep{2012MNRAS.420.1462R, 2019MNRAS.484.2711R}, that are possibly a product of non--canonical formation channels, as close binary evolution and mergers.

\section*{Acknowledgements}

We thank the anonymous referee for the useful comments and suggestions.
ADR and SOK acknowledges financial support from CNPq and PRONEX-FAPERGS/CNPq (Brazil). LAA and LF acknowledges financial support from CNPq (Brazil).
Based on observations at the Southern Astrophysical Research (SOAR) telescope, which is a joint project of MCTIC--Brasil, US--NOAO, the University of North Carolina at Chapel Hill (UNC), and Michigan State University (MSU) and processed using the IRAF package, developed by the Association of Universities for Research in Astronomy, Inc. This paper includes data collected with the TESS mission, obtained from the MAST data archive at the Space Telescope Science Institute (STScI). Funding for the TESS mission is provided by the NASA Explorer Program. 
This work has made use of data from the European Space Agency (ESA) mission Gaia (\url{https://www.cosmos.esa.int/gaia}), processed by the Gaia Data Processing and Analysis Consortium (DPAC, \url{https://www.cosmos.esa.int/web/gaia/dpac/consortium}). Funding for the DPAC has been provided by national institutions, in particular the institutions participating in the Gaia Multilateral Agreement. This research  has also made use of the NASA Astrophysics Data System.

\bibliographystyle{mnras}
\bibliography{rev1_hot-DAV} 

\begin{thebibliography}{}
\makeatletter
\relax
\def\mn@urlcharsother{\let\do\@makeother \do\$\do\&\do\#\do\^\do\_\do\%\do\~}
\def\mn@doi{\begingroup\mn@urlcharsother \@ifnextchar [ {\mn@doi@}
  {\mn@doi@[]}}
\def\mn@doi@[#1]#2{\def\@tempa{#1}\ifx\@tempa\@empty \href
  {http://dx.doi.org/#2} {doi:#2}\else \href {http://dx.doi.org/#2} {#1}\fi
  \endgroup}
\def\mn@eprint#1#2{\mn@eprint@#1:#2::\@nil}
\def\mn@eprint@arXiv#1{\href {http://arxiv.org/abs/#1} {{\tt arXiv:#1}}}
\def\mn@eprint@dblp#1{\href {http://dblp.uni-trier.de/rec/bibtex/#1.xml}
  {dblp:#1}}
\def\mn@eprint@#1:#2:#3:#4\@nil{\def\@tempa {#1}\def\@tempb {#2}\def\@tempc
  {#3}\ifx \@tempc \@empty \let \@tempc \@tempb \let \@tempb \@tempa \fi \ifx
  \@tempb \@empty \def\@tempb {arXiv}\fi \@ifundefined
  {mn@eprint@\@tempb}{\@tempb:\@tempc}{\expandafter \expandafter \csname
  mn@eprint@\@tempb\endcsname \expandafter{\@tempc}}}

\bibitem[\protect\citeauthoryear{{Althaus}, {Panei}, {Miller Bertolami},
  {Garc{\'\i}a-Berro}, {C{\'o}rsico}, {Romero}, {Kepler}  \&
  {Rohrmann}}{{Althaus} et~al.}{2009}]{2009ApJ...704.1605A}
{Althaus} L.~G.,  {Panei} J.~A.,  {Miller Bertolami} M.~M.,
  {Garc{\'\i}a-Berro} E.,  {C{\'o}rsico} A.~H.,  {Romero} A.~D.,  {Kepler}
  S.~O.,   {Rohrmann} R.~D.,  2009, \mn@doi [\apj]
  {10.1088/0004-637X/704/2/1605}, \href
  {https://ui.adsabs.harvard.edu/abs/2009ApJ...704.1605A} {704, 1605}

\bibitem[\protect\citeauthoryear{{Bergeron}, {Wesemael}  \&
  {Beauchamp}}{{Bergeron} et~al.}{1995}]{1995PASP..107.1047B}
{Bergeron} P.,  {Wesemael} F.,   {Beauchamp} A.,  1995, \mn@doi [\pasp]
  {10.1086/133661}, \href
  {https://ui.adsabs.harvard.edu/abs/1995PASP..107.1047B} {107, 1047}

\bibitem[\protect\citeauthoryear{{Bogn{\'a}r} et~al.,}{{Bogn{\'a}r}
  et~al.}{2020}]{2020arXiv200311481B}
{Bogn{\'a}r} Z.,  et~al., 2020, arXiv e-prints, \href
  {https://ui.adsabs.harvard.edu/abs/2020arXiv200311481B} {p. arXiv:2003.11481}

\bibitem[\protect\citeauthoryear{{Fontaine} \& {Wesemael}}{{Fontaine} \&
  {Wesemael}}{1987}]{1987fbs..conf..319F}
{Fontaine} G.,  {Wesemael} F.,  1987, in {Philip} A.~G.~D.,  {Hayes} D.~S.,
  {Liebert} J.~W.,  eds, IAU Colloq. 95: Second Conference on Faint Blue Stars.
  pp 319--326

\bibitem[\protect\citeauthoryear{{Gianninas}, {Bergeron}  \&
  {Ruiz}}{{Gianninas} et~al.}{2011}]{2011ApJ...743..138G}
{Gianninas} A.,  {Bergeron} P.,   {Ruiz} M.~T.,  2011, \mn@doi [\apj]
  {10.1088/0004-637X/743/2/138}, \href
  {https://ui.adsabs.harvard.edu/abs/2011ApJ...743..138G} {743, 138}

\bibitem[\protect\citeauthoryear{{Kepler} et~al.,}{{Kepler}
  et~al.}{2016}]{2016MNRAS.455.3413K}
{Kepler} S.~O.,  et~al., 2016, \mn@doi [\mnras] {10.1093/mnras/stv2526}, \href
  {https://ui.adsabs.harvard.edu/abs/2016MNRAS.455.3413K} {455, 3413}

\bibitem[\protect\citeauthoryear{{Kepler}, {Romero}, {Pelisoli}  \&
  {Ourique}}{{Kepler} et~al.}{2017}]{2017IJMPS..4560023K}
{Kepler} S.~O.,  {Romero} A.~D.,  {Pelisoli} I.,   {Ourique} G.,  2017, in
  International Journal of Modern Physics Conference Series. p. 1760023
  (\mn@eprint {arXiv} {1702.01134}), \mn@doi{10.1142/S2010194517600230}

\bibitem[\protect\citeauthoryear{{Koester} \& {Kepler}}{{Koester} \&
  {Kepler}}{2015}]{2015A&A...583A..86K}
{Koester} D.,  {Kepler} S.~O.,  2015, \mn@doi [\aap]
  {10.1051/0004-6361/201527169}, \href
  {https://ui.adsabs.harvard.edu/abs/2015A&A...583A..86K} {583, A86}

\bibitem[\protect\citeauthoryear{{Koester}, {Voss}, {Napiwotzki}, {Christlieb},
  {Homeier}, {Lisker}, {Reimers}  \& {Heber}}{{Koester}
  et~al.}{2009}]{2009A&A...505..441K}
{Koester} D.,  {Voss} B.,  {Napiwotzki} R.,  {Christlieb} N.,  {Homeier} D.,
  {Lisker} T.,  {Reimers} D.,   {Heber} U.,  2009, \mn@doi [\aap]
  {10.1051/0004-6361/200912531}, \href
  {https://ui.adsabs.harvard.edu/abs/2009A&A...505..441K} {505, 441}

\bibitem[\protect\citeauthoryear{{Kurtz}, {Shibahashi}, {Dhillon}, {Marsh}  \&
  {Littlefair}}{{Kurtz} et~al.}{2008}]{Kurtz2008}
{Kurtz} D.~W.,  {Shibahashi} H.,  {Dhillon} V.~S.,  {Marsh} T.~R.,
  {Littlefair} S.~P.,  2008, \mn@doi [\mnras]
  {10.1111/j.1365-2966.2008.13664.x}, \href
  {https://ui.adsabs.harvard.edu/abs/2008MNRAS.389.1771K} {389, 1771}

\bibitem[\protect\citeauthoryear{{Kurtz}, {Shibahashi}, {Dhillon}, {Marsh},
  {Littlefair}, {Copperwheat}, {G{\"a}nsicke}  \& {Parsons}}{{Kurtz}
  et~al.}{2013}]{Kurtz2013}
{Kurtz} D.~W.,  {Shibahashi} H.,  {Dhillon} V.~S.,  {Marsh} T.~R.,
  {Littlefair} S.~P.,  {Copperwheat} C.~M.,  {G{\"a}nsicke} B.~T.,   {Parsons}
  S.~G.,  2013, \mn@doi [\mnras] {10.1093/mnras/stt585}, \href
  {https://ui.adsabs.harvard.edu/abs/2013MNRAS.432.1632K} {432, 1632}

\bibitem[\protect\citeauthoryear{{Kurtz}, {Shibahashi}, {Murphy}, {Bedding}  \&
  {Bowman}}{{Kurtz} et~al.}{2015}]{2015MNRAS.450.3015K}
{Kurtz} D.~W.,  {Shibahashi} H.,  {Murphy} S.~J.,  {Bedding} T.~R.,   {Bowman}
  D.~M.,  2015, \mn@doi [\mnras] {10.1093/mnras/stv868}, \href
  {https://ui.adsabs.harvard.edu/abs/2015MNRAS.450.3015K} {450, 3015}

\bibitem[\protect\citeauthoryear{{Lenz} \& {Breger}}{{Lenz} \&
  {Breger}}{2004}]{2004IAUS..224..786L}
{Lenz} P.,  {Breger} M.,  2004, in {Zverko} J.,  {Ziznovsky} J.,  {Adelman}
  S.~J.,   {Weiss} W.~W.,  eds,  IAU Symposium Vol. 224, The A-Star Puzzle. pp
  786--790, \mn@doi{10.1017/S1743921305009750}

\bibitem[\protect\citeauthoryear{{Robinson}, {Kepler}  \& {Nather}}{{Robinson}
  et~al.}{1982}]{RKN}
{Robinson} E.~L.,  {Kepler} S.~O.,   {Nather} R.~E.,  1982, \mn@doi [\apj]
  {10.1086/160162}, \href
  {https://ui.adsabs.harvard.edu/abs/1982ApJ...259..219R} {259, 219}

\bibitem[\protect\citeauthoryear{{Romero} \& {Istrate}}{{Romero} \&
  {Istrate}}{2019}]{2019cwdb.confE..13R}
{Romero} A.~D.,  {Istrate} A.~G.,  2019, in {Tovmassian} G.~H.,  {Gansicke}
  B.~T.,  eds, Compact White Dwarf Binaries. p.~13

\bibitem[\protect\citeauthoryear{{Romero}, {C{\'o}rsico}, {Althaus}, {Kepler},
  {Castanheira}  \& {Miller Bertolami}}{{Romero}
  et~al.}{2012}]{2012MNRAS.420.1462R}
{Romero} A.~D.,  {C{\'o}rsico} A.~H.,  {Althaus} L.~G.,  {Kepler} S.~O.,
  {Castanheira} B.~G.,   {Miller Bertolami} M.~M.,  2012, \mn@doi [\mnras]
  {10.1111/j.1365-2966.2011.20134.x}, \href
  {https://ui.adsabs.harvard.edu/abs/2012MNRAS.420.1462R} {420, 1462}

\bibitem[\protect\citeauthoryear{{Romero}, {Kepler}, {Joyce}, {Lauffer}  \&
  {C{\'o}rsico}}{{Romero} et~al.}{2019a}]{2019MNRAS.484.2711R}
{Romero} A.~D.,  {Kepler} S.~O.,  {Joyce} S.~R.~G.,  {Lauffer} G.~R.,
  {C{\'o}rsico} A.~H.,  2019a, \mn@doi [\mnras] {10.1093/mnras/stz160}, \href
  {https://ui.adsabs.harvard.edu/abs/2019MNRAS.484.2711R} {484, 2711}

\bibitem[\protect\citeauthoryear{{Romero} et~al.,}{{Romero}
  et~al.}{2019b}]{2019MNRAS.490.1803R}
{Romero} A.~D.,  et~al., 2019b, \mn@doi [\mnras] {10.1093/mnras/stz2571}, \href
  {https://ui.adsabs.harvard.edu/abs/2019MNRAS.490.1803R} {490, 1803}

\bibitem[\protect\citeauthoryear{{Shibahashi}}{{Shibahashi}}{2005}]{2005EAS....17..143S}
{Shibahashi} H.,  2005, in {Alecian} G.,  {Richard} O.,   {Vauclair} S.,  eds,
  EAS Publications Series Vol. 17, EAS Publications Series. pp 143--148,
  \mn@doi{10.1051/eas:2005108}

\bibitem[\protect\citeauthoryear{{Shibahashi}}{{Shibahashi}}{2007}]{2007AIPC..948...35S}
{Shibahashi} H.,  2007, in {Stancliffe} R.~J.,  {Houdek} G.,  {Martin} R.~G.,
  {Tout} C.~A.,  eds,  American Institute of Physics Conference Series Vol.
  948, Unsolved Problems in Stellar Physics: A Conference in Honor of Douglas
  Gough. pp 35--42, \mn@doi{10.1063/1.2818994}

\bibitem[\protect\citeauthoryear{{Tassoul}}{{Tassoul}}{1980}]{1980ApJS...43..469T}
{Tassoul} M.,  1980, \mn@doi [\apjs] {10.1086/190678}, \href
  {https://ui.adsabs.harvard.edu/abs/1980ApJS...43..469T} {43, 469}

\bibitem[\protect\citeauthoryear{{Vanderbosch}, {Winget}  \&
  {Winget}}{{Vanderbosch} et~al.}{2018}]{Zach2018}
{Vanderbosch} Z.~P.,  {Winget} K.~I.,   {Winget} D.~E.,  2018, in 21th European
  White Dwarf Workshop, Austin, Texas.

\bibitem[\protect\citeauthoryear{{Winget}, {van Horn}, {Tassoul}, {Fontaine},
  {Hansen}  \& {Carroll}}{{Winget} et~al.}{1982}]{1982ApJ...252L..65W}
{Winget} D.~E.,  {van Horn} H.~M.,  {Tassoul} M.,  {Fontaine} G.,  {Hansen}
  C.~J.,   {Carroll} B.~W.,  1982, \mn@doi [\apjl] {10.1086/183721}, \href
  {https://ui.adsabs.harvard.edu/abs/1982ApJ...252L..65W} {252, L65}

\makeatother
\end{thebibliography}

\end{document}